\documentclass[12pt,a4paper]{article}
\usepackage{rotating}
\usepackage{authblk}
\usepackage[top=2cm, bottom=2cm, left=2cm, right=2cm, footnotesep=1cm]{geometry}
\usepackage[round]{natbib}
\usepackage{graphicx}
\usepackage{subcaption}
\usepackage{tabulary}
\usepackage{multirow}
\usepackage{setspace}
\usepackage{amsmath}
\usepackage{amsthm}
\usepackage{amssymb}
\usepackage{multirow}
\usepackage[titletoc,title]{appendix}
\usepackage[colorlinks,linkcolor=black,anchorcolor=black,citecolor=black]{hyperref}
\usepackage{lineno}
\usepackage[shortlabels]{enumitem}
\usepackage{tikz}
\usepackage{siunitx}
\usepackage{ulem}
\usepackage{xfrac}
\usepackage{url}
\usepackage{palatino}
\usepackage{newpxtext}
\usepackage{newpxmath}

\DeclareMathOperator*{\argmax}{arg\,max}

\hypersetup{
    colorlinks = {false},
    allbordercolors = {white},
}

\let\Contentsline\contentsline
\renewcommand

\newtheorem{property}{Property}

\newtheoremstyle{property}% name
{3pt}% space above
{3pt}% space below
{}% body font
{}% indent amount
{}% theorem head font
{.}% punctuation after theorem head
{.5em}% space after theorem head
{}% theorem head spec
\newtheoremstyle{observation}% name
{3pt}% space above
{3pt}% space below
{}% body font
{}% indent amount
{}% theorem head font
{.}% punctuation after theorem head
{.5em}% space after theorem head
{}% theorem head spec

\title{A vicious cycle along busy bus corridors and how to abate it}

\author[a,b]{Minyu Shen}
\author[b]{Weihua Gu\footnote{Corresponding author, Email address:\,\href{mailto:weihua.gu@polyu.edu.hk}{weihua.gu@polyu.edu.hk}}}
\author[c]{Michael J. Cassidy}
\author[d]{Yongjie Lin}
\author[c]{Wei Ni}

\affil[a]{School of Management Science and Engineering, Southwestern University of Finance and Economics, China} 
\affil[b]{Department of Electrical Engineering, The Hong Kong Polytechnic University}
\affil[c]{Department of Civil and Environmental Engineering, University of California, Berkeley}
\affil[d]{School of Civil Engineering and Transportation, South China University of Technology}
\date{\vspace{-2.8em}}

\begin{document}
\maketitle
\noindent\hrulefill
\vspace{-1.2em}

%\linenumbers
%\begin{pagewiselinenumbers}
\section*{Abstract}
We unveil that a previously-unreported vicious cycle can be created when bus queues form at curbside stops along a corridor. Buses caught in this cycle exhibit growing variation in headways as they travel from stop to stop. Bus (and patron) delays accumulate in like fashion and can grow large on long, busy corridors. We show that this damaging cycle can be abated in simple ways. Present solutions entail holding buses at a corridor entrance and releasing them as per various strategies proposed in the literature. We introduce a modest variant to the simplest of these strategies. It releases buses at headways that are slightly less than, or equal to, the scheduled values. It turns out that periodically releasing buses at slightly smaller headways can substantially reduce bus delays caused by holding so that benefits can more readily outweigh costs in corridors that contain a sufficient number of serial bus stops. The simple variant is shown to perform about as well as, or better than, other bus-holding strategies in terms of saving delays, and is more effective than other strategies in regularizing bus headways. We also show that grouping buses from across multiple lines and holding them by group can be effective when patrons have the flexibility to choose buses from across all lines in a group. Findings come by formulating select models of bus-corridor dynamics and using these to simulate part of the Bus Rapid Transit corridor in Guangzhou, China.\\\\
\textbf{Keywords:} bus corridor; bus queues; bus holding; common-line patrons

\noindent\hrulefill

\section{Introduction}\label{sec:introduction}
%Examples include: the BRS Presidente Vargas Corridor in Rio de Janeiro, Brazil, where peak bus flows are 600 per hour in each direction; and the Bus Rapid Transit (BRT) corridor in Guangzhou, China, which serves patron flows as high as 27,000/h/direction during the rush \citep{brtdata}.

Busy bus corridors are found the world over. Examples include: the Santo Amaro-Nove de Julho-Centro corridor in Sao Paulo, Brazil, which serves patron flows as high as 98,000/h/\break direction during peak hours; and the Bus Rapid Transit (BRT) corridor in Guangzhou, China, accommodating patron flows of 27,000/h/direction during the rush \citep{brtdata}. Bus queues that form at stops in corridors such as these create irregular headways. Buses with longer headways can confront larger numbers of boarding patrons at downstream stops. These buses are further delayed as a result, which causes other buses to catch-up from behind, a phenomenon called bunching \citep{newell1964maintaining}. \par

Bunching is known to become more pronounced as buses progress along their routes \citep{newell1974control, daganzo2009headway}. One might therefore expect attendant variations in bus headways to worsen in this same fashion. And because large variations in headways engender long queues (see \citealp{gu2011capacity}), one might further expect to see steadily longer bus queues when moving from stop to stop along a corridor. This would mean that schedule-keeping erodes, and delays to buses and patrons increase, as buses traverse the corridor. This vicious cycle, triggered by bus queues at stops could also worsen over the course of a rush.\par

The plausibility of the cycle described above seems to have escaped notice in the literature. Perhaps this is because studies of bus queueing often focus on a single stop, rather than on multi-stop corridors \citep{gibson1989bus, fernandez2002capacity, fernandez2005effect, fernandez2010modelling, gu2011capacity, gu2015models, gu2013maximizing, tan2014berth,bian2019performance, shen2019capacity, shen2023efficient}. And those works to have examined corridor-wide operations commonly assumed that stops were free of bus queues \citep{hernandez2015analysis, schmocker2016bus, laskaris2018multiline}.\par

The literature also reports on several forms of control that were examined for idealized (unqueued) corridor-wide conditions. These forms include adaptive signal control \citep{anderson2020effect}; stop-skipping \citep{fu2003real, sun2005real}; and the imposition of boarding limits \citep{delgado2009real}. The most widely-studied control form is bus holding \citep{osuna1972control,vandebona1986effect, hickman2001analytic,daganzo2009headway,xuan2011dynamic,wu2017modelling,berrebi2018comparing}. The latter form regularizes headways by holding buses at select stops, termed control points, along a corridor. Some of the earliest efforts in this realm entailed adding fixed slack times to bus schedules and prohibiting buses from departing control points ahead of the new schedule \citep{osuna1972control,newell1974control,zhao2006optimal}. Work in \citet{daganzo2009headway} proposed a dynamic strategy in which holding times are determined at multiple control points along a line based upon bus headways measured in real time. The work was extended in numerous follow-up studies \citep{daganzo2011reducing,xuan2011dynamic,bartholdi2012self,he2015anti,andres2017predictive,zhang2018two,dai2019predictive} to formulate more sophisticated holding strategies that achieve better performance. Finally, work in \citet{berrebi2015real} found bus-holding times that minimize headway variations based upon predictions of bus arrival times in the future.\par

All the above holding strategies enhance schedule-keeping. However, in a corridor free of bus queues (as assumed in the above works), these strategies can only slow buses down and thus worsen delay for both buses and patrons, as compared against doing nothing. The strategies can reduce delays, but only when buses queue-up at stops, and these cases were not considered in the above works.\par

We found three works that considered bus queues in multi-stop corridors, though none of these are aligned with our present interest. The first of these studies examined patron behavior in choosing among multiple lines that serve their trips \citep{sun2018considering}. Another focused on reducing patron wait times by controlling bus speeds \citep{bian2020optimization}. The third work sought to mitigate bus-stop queues by coordinating the cruising speeds of buses across different lines to distribute bus arrivals at stops more evenly over time \citep{bian2023real}. None of these studies mentioned the vicious cycle described earlier in this section, much less strategies to mitigate it.\par

The present paper formulates models of corridor dynamics and uses them in simulation to: (i) verify that the vicious cycle can occur in commonplace settings; (ii) illustrate its damaging effects; and (iii) show that the cycle can be abated by holding buses at a single control point at the corridor's entrance, to reduce their headway variations and as a result, their delays. Holding strategies to be tested include small variations in the simple headway-based strategy studied in both \citet{abkowitz1990implementing} and \citet{berrebi2018comparing}. In our simple variants, buses are released from the control point at regular headways that can be slightly shorter than scheduled values and that can be for bus groupings across distinct lines. The variant is shown to be effective, often more so than more complex holding strategies found in the literature.\par

The idealized features of a bus corridor, its control point, and the ideas behind the simple headway-based holding strategy are described in the following section. Models of corridor operations and what these say about how the simple headway-based holding strategy might be favorably altered are presented in Section~\ref{sec:model_logic}. The pulling-together of these models into a simulation platform and the real-world site that provided input data to our analysis are described in Section~\ref{sec:simulation}. Section~\ref{sec:numerical_analysis} provides outcomes from numerical tests on the simple strategy, where buses are released when their headways meet the scheduled value, and on its variant that employs a reduced headway threshold for releasing buses onto the corridor. A second variant of the simple strategy that holds buses by line group, along with its benefits, are presented in Section~\ref{sec:common_line}. Implications of outcomes and their  extensions are discussed in Section~\ref{sec:conclusion}.

%Notation is defined as needed throughout the paper, and is summarized in Appendix \ref{apdx:notation} for the reader's convenience.

\section{Corridor Representation and Holding}\label{sec:setup}

\begin{figure}[h!]
    \centering
    \includegraphics[scale=0.5]{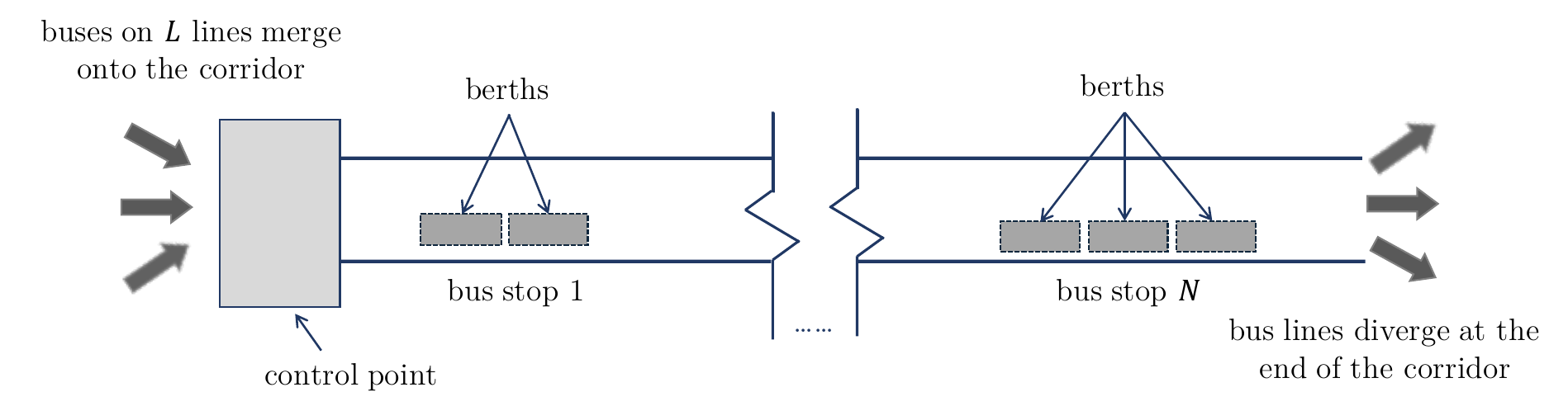}
    \caption{Corridor layout.}\label{fig:corridor}
\end{figure}

Consider a corridor serving $L$ bus lines, as shown in Figure~\ref{fig:corridor}. Lines may originate at distinct locations upstream and merge onto the corridor. The lines run through each of $N$ multi-berth stops that reside curbside along the corridor, and diverge downstream. Denote $c^s$ as the number of berths at stop $s\in\{1,2,\ldots,N\}$, and assume that buses are not allowed to overtake one another when they enter and exit these berths at any stop.\footnote{Other queue disciplines, such as limited- and free-overtaking \citep{gu2013maximizing,bian2019performance,hu2023impacts}, can be incorporated into our models and would not likely alter present findings as regards the existence of a vicious cycle and the effectiveness of bus holding as a mitigation measure.} Denote as $f_l$ the bus flow (i.e., service frequency) on line $l\in\{1,2,...,L\}$. The bus headway scheduled for line $l$, $H_l$, is therefore $\frac{1}{f_l}$.

\par

\begin{figure}[h!]
    \centering
    \includegraphics[scale=0.5]{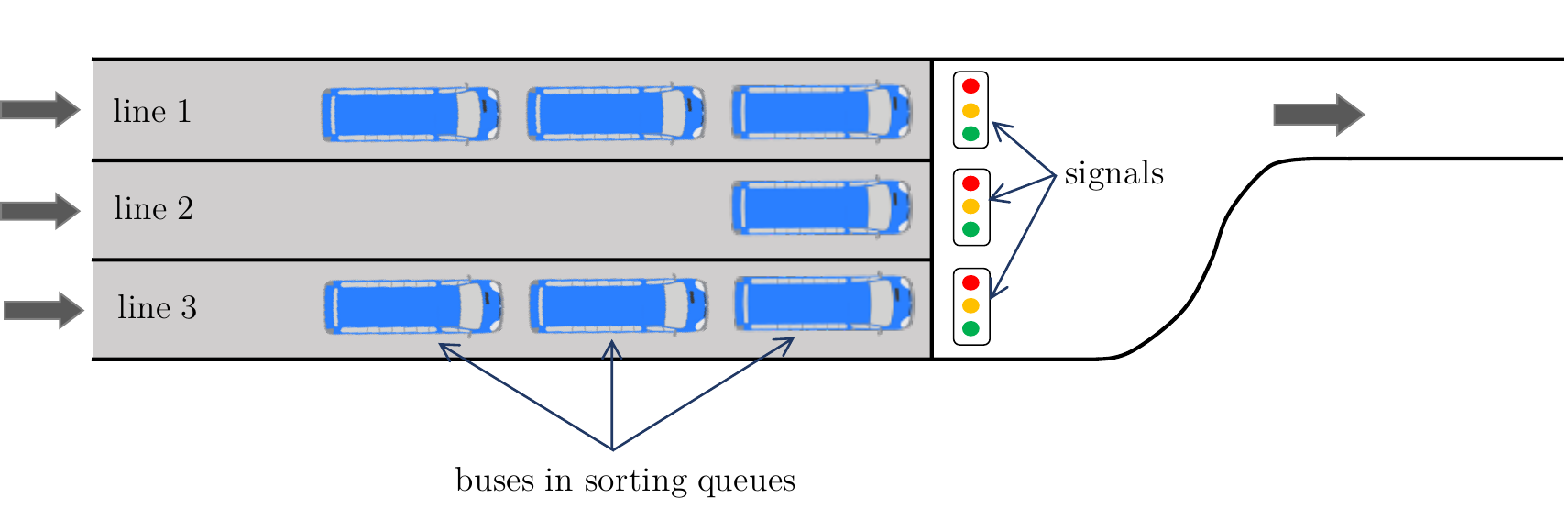}
    \caption{An example layout of a control point ($L=3$).}\label{fig:sorting_point_by_line}
\end{figure}

Buses are held and released at the control point located at the corridor's upstream end, again as shown in Figure~\ref{fig:corridor}. Figure~\ref{fig:sorting_point_by_line} illustrates how this control point might be designed for a case of $L=3$, where $L$ dedicated bus lanes are used, each serving one of $L$ bus lines. (A similar design is offered in \citealp{szasz1978comonor}.) Buses arriving on line $l$ might first queue in the lane assigned to that line. A bus is released from the control point at a time $H_l$ following the release of the bus ahead. A line-$l$ bus arriving at an empty lane is released into the corridor immediately, if the elapsed time since the line's previous release has reached or exceeded $H_l$. Bus releases might be controlled by signals (meters) installed at the downstream ends of reserved lanes; see again Figure~\ref{fig:sorting_point_by_line}.\par

 \par

Where space is at a premium, bus queues from distinct lanes can be arranged in tandem fashion in a single reserved lane. If real-time bus-to-bus communication is available as described in \citet{argote2015dynamic}, then control can be executed by regulating bus speeds, rather than by holding buses, thus eliminating need for a control point.\par

Where holding is the means of control, a bus $j$ on line $l$ encounters a holding delay, denoted $w_{l,j}^0$. The control method therefore reduces bus delay on the corridor only if holding delays are outsized by the accumulated bus delays that holding saves at downstream stops. (This can only be the case when stops are queued with buses sans holding.)\par

We will show that properties of holding delays can in some circumstances motivate modifications to the simple headway-based holding strategy described above. In one modification, buses on line $l$ are released from the control point at intervals that are slightly less than $H_l$. A second modification entails grouping buses from across lines and holding them at the control point by group. These matters are discussed in the following sections.

\section{Models}\label{sec:model_logic}

Models of bus dynamics and of patron boardings and alightings are presented in Sections \ref{sec:bus_dynamics} and \ref{sec:boarding_dynamic}, respectively. Bus-delay equations for assessing corridor performance are in Section \ref{sec:delay_metric}.
\subsection{Bus dynamics}\label{sec:bus_dynamics}
Assume for simplicity that bus travel time between the control point and first stop downstream (i.e., stop 1 in Figure~\ref{fig:corridor}) is zero. Hence, the arrival time at that stop by bus $j$ on line $l$, $a_{l,j}^1$, is calculated as:

\begin{equation}\label{eq:first_stop_arrive_time}
	a_{l,j}^1=a_{l,j}^0+w_{l,j}^0,l\in\{1,2,...,L\}, j\in\{1,2,3,...\},
\end{equation}
where $a_{l,j}^0$ is bus $j$'s arrival time at the control point, and recall that $w_{l,j}^0$ is the bus's holding delay. Assume that $a_{l,j}^0$ is a Gaussian random variable with mean $jH_l$ and standard deviation $C_{H,l}H_l$, i.e., $a_{l,j}^0 \sim \mathcal{N}(jH_l, C_{H,l}^2 H_l^2)$, where $C_{H,l}$ is a specified coefficient.\footnote{The coefficient of variation of headways is $\sqrt{2}C_{H,l}$ if the bus arrival times are independent \citep{ross2020introduction}. Furthermore, our simulation tests found that other distributions for $a_{l,j}^0$ produced similar outcomes.} In cases in which bus overtaking occurs between stops, such that $a_{l,j}^0 > a_{l,j+1}^0$, the indices of the two buses are swapped.\par

The expected holding delay for bus $j$ on line $l$ can be approximated as:

\begin{equation}\label{eq:approx_delay}
	E[w^0_{l,j}|j] \approx C_{H,l}H_l\cdot \Phi^{-1}\left( \frac{j-\frac{\pi}{8}}{j-\frac{\pi}{4}+1} \right),
\end{equation}
where $\Phi^{-1}(\cdot)$ is the inverse cumulative distribution function for the standard normal distribution. Derivation of (\ref{eq:approx_delay}) is relegated to Appendix \ref{apdx:approx_delay}.\par

In light of (\ref{eq:approx_delay}), average holding delay across the first $m$ buses on line $l$ is: 
\begin{equation}\label{eq:first_m_avg_delay}
	E_{1\leq j\leq m}[w_{l,j}^0] \approx C_{H,l}H_l\cdot \frac{1}{m}\sum_{j=1}^{m}\Phi^{-1}\left( \frac{j-\frac{\pi}{8}}{j-\frac{\pi}{4}+1} \right).
\end{equation}

Equations (\ref{eq:approx_delay}) and (\ref{eq:first_m_avg_delay}) unveil the following two properties regarding holding delays.

\begin{property}\label{property:1}\normalfont
	Expected holding delay is proportional to the standard deviation of bus arrival time, $C_{H,l}H_l$.
\end{property}
\begin{property}\label{property:2}\normalfont
	The expected delay for bus $j$, $E[w_{l,j}^0|j]$,  grows to infinity as $j$ increases, since $\frac{j-\frac{\pi}{8}}{j-\frac{\pi}{4}+1}$ increases to 1 as $j$ approaches infinity. In addition, since $\{ E[w_{l,j}^0|j] \}$ is a divergent increasing sequence, the sequence of its cumulative means, $\{ E_{1\leq j\leq m}[w_{l,j}^0] \}$, is also a divergent increasing sequence. (Proof of this mathematical result is provided in Appendix \ref{apdx:abel_proof}.) Thus, average holding delay (\ref{eq:first_m_avg_delay}) also approaches infinity as $m$ grows. Fortunately, these growths toward infinity are slow.\footnote{Numerical analysis shows that the coefficient in (\ref{eq:approx_delay}), $\Phi^{-1}\left( \frac{j-\frac{\pi}{8}}{j-\frac{\pi}{4}+1} \right)$, moderately increases from 0.60 to 2.87 as $j$ increases from 2 to 300. The coefficient in (\ref{eq:first_m_avg_delay}), $\frac{1}{m}\sum_{j=1}^{m}\Phi^{-1}\left( \frac{j-\frac{\pi}{8}}{j-\frac{\pi}{4}+1} \right)$, grows even slower. Since the number of buses entering a corridor during a peak period is in the hundreds or less, the holding delay will not grow exceedingly large.}
\end{property}

Property \ref{property:2} means that under a simple headway-based control scheme where buses are released from the control point at scheduled headways, those buses will fall further and further behind schedule as the rush wears on. Both to combat this tendency and to reduce holding delay, we propose the following modification of the simple headway-based strategy studied in \citet{abkowitz1990implementing} and \citet{berrebi2018comparing}. With our modification, buses on each line $l\in\{1,2,\ldots,L\}$ are held at the control point until a minimum headway of $\eta H_l$ elapses between consecutive departures from the same line, where $\eta\in [0,1]$.\par

Previous studies (e.g., \citealp{daganzo2009headway}) often model the uncertainties in bus motion as exogenous random terms. However, the stochastic delay incurred by a queued bus at a busy stop is affected by downstream buses in the queue, and thus should be treated endogenously. This bus-queueing dynamic is modeled as described below.\par

Since bus overtaking maneuvers are prohibited inside the multi-berth stops, an arriving bus must enter a queue if the stop's upstream-most berth is occupied. The delay imposed by the queue at stop $s$ is denoted $q_{l,j}^s$. Assume that these queues are not affected by queues from other bottlenecks nearby. When a bus finishes loading and unloading its patrons, it may be blocked by another bus dwelling in a downstream berth in the same stop, and thus incurs an in-berth delay, $b_{l,j}^s$. The bus's departure time from the stop is therefore:

\begin{equation}
	d_{l,j}^s=a_{l,j}^s+q_{l,j}^s+S_{l,j}^s+b_{l,j}^s,l\in\{1,2,...,L\},j\in\{1,2,3,...\},s\in\{1,2,...,N\},
\end{equation}
where $d_{l,j}^s$ and $a_{l,j}^s$ denote the bus's departure and arrival times from and to the stop, respectively; $S_{l,j}^s$ the bus's dwell time in loading and unloading patrons at the stop; and delay terms $q_{l,j}^s$ and $b_{l,j}^s$ will be determined dynamically through simulation.\par

A bus's arrival time at downstream stop $s+1$, denoted $a_{l,j}^{s+1}$, is determined as:\begin{equation}
	a_{l,j}^{s+1} = d_{l,j}^s + t_{l,j}^s, l\in\{1,2,...,L\},j\in\{1,2,3,...\},s\in\{1,2,...,N\},
\end{equation}
where $t_{l,j}^s$ is the bus's inter-stop travel time from stop $s$ to stop $s+1$. The $t_{l,j}^s$ is assumed to follow a lognormal distribution with mean $\mu_{T}^s$ and standard deviation $\sigma_{T}^{s}$; i.e., $t_{l,j}^s \sim \mathcal{LN} (\mu_{T}^s, (\sigma_{T}^{s})^2)$, as per empirical findings in \citet{kieu2014establishing}.\footnote{This implies that buses can overtake each other when traveling between stops, which may not be allowed, particularly among buses on the same line. Yet, our simulation tests show that use of the lognormal distribution provides good approximations even when inter-stop overtaking maneuvers are prohibited.} The $\sigma_{T}^{s}$ reflects the extent to which buses are disturbed while traveling between stops. A lower $\sigma_{T}^{s}$ may occur when a dedicated bus lane or signal priority is provided.\par

\subsection{Bus dwell time model and common-line patrons}\label{sec:boarding_dynamic}

Assume that bus dwell time, $S_{l,j}^s$, is dictated by a linear function of the numbers of boarding and alighting patrons, $p_{l,j}^s$ and $h_{l,j}^s$, such that
\begin{equation}\label{eq:service_pax}
	S_{l,j}^{s} = \tau + \delta_{b}^{s}p_{l,j}^{s} + \delta_{a}^{s}h_{l,j}^s, l\in \{1,2,...,L\}, j\in \{1,2,3,...\}, s\in \{1,2,...,N\},
\end{equation}
where $\tau$ is the time lost due to bus deceleration and acceleration at a stop, and $\delta_b^s$ and $\delta_a^s$ are the boarding and alighting times per patron at stop $s$, respectively.\par

In estimating $p_{l,j}^s$, we note that some patrons may be able to choose buses from among multiple lines \citep{cominetti2001common,schmocker2016bus, laskaris2018multiline}. Common-line trips of this sort are modeled by dividing the $L$ bus lines into $K\leq L$ line groups. We assume that patrons can also be divided into $K$ mutually-exclusive groups, so that patrons in one group will not board buses from other groups.\footnote{In practice, a suitable grouping can be determined knowing: the OD pairs of all patrons, and the set of lines that can serve each OD pair. Lines that share common-line trip-making can be grouped together.} Common-line patrons within the $k$-th line group ($k\in \{1,2,\ldots,K\}$) are assumed to arrive at stop $s$ at rate $\lambda_{k,C}^s$. Those patrons can board buses on any line in that group. Non-common-line patrons on line $l$ arrive at stop $s$ at rate $\lambda_{l}^s$, and can only ride buses on that line.\par

Buses are assumed to have sufficient onboard capacity to admit all patrons wishing to board, including those who arrive while the bus is dwelling \citep{osuna1972control, schmocker2016bus}.\footnote{This assumption may overestimate bus bunching and queueing since in reality, delayed and crowded buses may admit few patrons and therefore incur shorter dwell times at downstream stops \citep{estrada2016bus}. In contrast, the counterpart may underestimate bunching and queueing due to the following two reasons. First, dwell times are more varied when crowded buses take few boarding patrons due to limited capacity. This will lead to larger queueing delays \citep{gu2011capacity,gu2015models}. Second, as the number of onboard patrons approaches a bus's patron-carrying capacity, boarding time per patron becomes longer \citep{schmocker2016bus}, producing more severe queueing and bunching. We have performed simulations where buses have a finite patron-carrying capacity, and results show that the major findings presented in this paper still hold.} When multiple buses from the same line group are dwelling at a same stop, each common-line patron chooses the bus with the shortest boarding queue. The number of alighting patrons, $h_{l,j}^s$, can be estimated via simulation, knowing patron OD demand along all bus lines; see e.g., \citet{he2019approach}.

\subsection{Performance metrics}\label{sec:delay_metric}

Average bus delay at stop $s$, $w^s$, is given by:
\begin{equation}
	w^s=E_{l\in\{1,2,...,L\},j\in\{1,2,...\}} [d_{l,j}^s-a_{l,j}^s-S_{l,j}^s],s\in\{1,2,...,N\}.
\end{equation}\par

Average cumulative delay per bus upon departing from stop $s$ is:
\begin{equation}
	W^s=E_{l\in\{1,2,...,L\},j\in\{1,2,...\}} [w_{l,j}^0]+\sum_{i=1}^{s}{w^s},s\in\{1,2,...,N\}.
\end{equation}

\section{Simulation Models and Inputs}\label{sec:simulation}
A bus corridor with multiple stops in series is a queueing network \citep{shortle2018fundamentals}. If stops have multiple berths, as presently assumed, then each stop is a polling system that is itself difficult to solve (e.g., \citealp{takagi1988queuing}). The complex queueing dynamics described in Sections \ref{sec:bus_dynamics} and \ref{sec:boarding_dynamic} render exact analytical solutions to our bus-queueing problem even more difficult to formulate. Moreover, even if analytical models could be developed, they would be too complicated to be computationally efficient, and hence loose their advantage over simulation. The benefits of analytical models would therefore be small. In contrast, simulation is simpler, can emulate complex, real-world dynamics and can readily adapt to variations in the scenarios to be examined here. These features make simulation a better choice for the problem of present interest.\par

Simulation models presently available for bus operations were either developed for a single bus stop \citep{gibson1989bus,fernandez2010modelling}; failed to account for bus queues at stops \citep{wu2017modelling}; or required calibration of numerous parameters \citep{cortes2005microsimulation,cortes2007modeling}. The latter concern can make it difficult to evaluate wide-ranging input conditions, and can obscure factors that contribute most to the effectiveness of a control strategy. To its credit, a simulation logic offered in \citet{bian2020optimization} modeled a single bus line to a level of detail sufficient for the present effort. But that logic assumed that bus arrival patterns and dwell times could be expressed exogenously, and did not consider common-line patrons.\par
\begin{sidewaysfigure}
\includegraphics[scale=0.62]{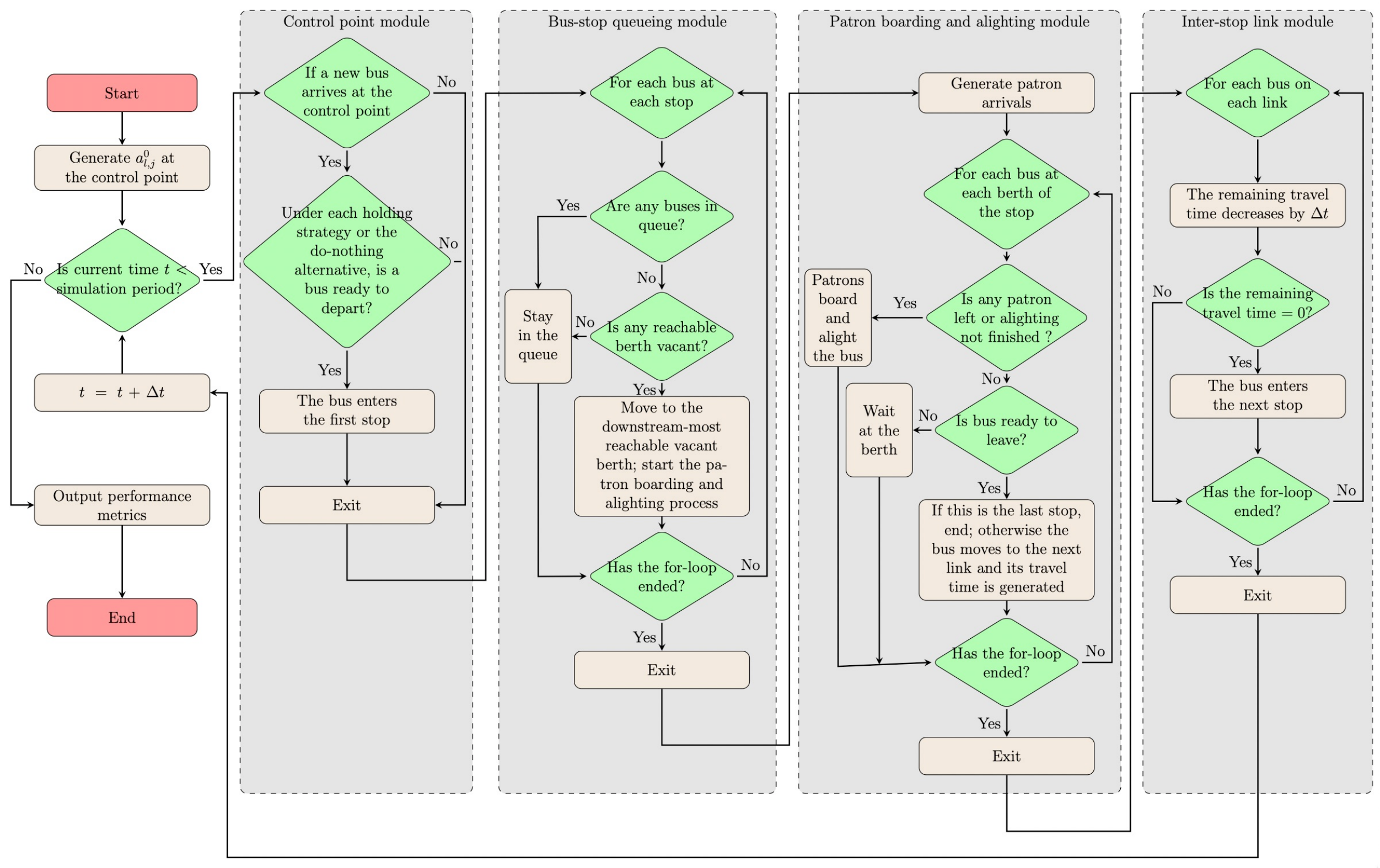}
\caption{Flowchart of simulation program.}\label{fig:flowchart}
\end{sidewaysfigure}

A parsimonious, discrete-time model of multi-line bus operations was therefore developed in-house and used for the present analysis. The model was coded in Python to emulate bus operations in four moduels: (1) at a control point; (2) at multi-berth stops; (3) as patrons board and alight; and (4) on links between stops. At each time step, the four modules are executed in sequence. Program logic is illustrated in Figure~\ref{fig:flowchart}. A visualization tool of bus motion was also developed. It uses simulation outputs as its inputs, and was used to check for programming errors. The code can be downloaded at: {https://github.com/Minyu-Shen/corrior-simulation.}

\subsection{Experimental procedures}
Each simulation started with a 1-hour warm-up time to mimic an off-peak period when bus queues might occur at some stops, but only sporadically. During this time, patron arrival rates at bus stops were set to only 30\% of the values specified for the rush to follow, and buses were not held at any control point. A 5-hour rush followed each warm-up period, during which time buses were held at the upstream end of the corridor and patron arrival rates at bus stops were based on observations described in Section~\ref{sec:data_collection}. \par

Simulations were repeated sufficient numbers of times to ensure that the performance metrics of Section \ref{sec:delay_metric} converged. The number of repetitions was selected as per \citet{sheldon2014prob} for each set of inputs, so that the estimated variance of $w^s$ did not exceed \num{5e-4} $\text{min}^2$. All simulation tests were performed on a Mac mini 2020 with M1 processor and 16 GB DDR4 memory. Each simulation required 1.5\,s, on average, to complete.

\subsection{Study site}\label{sec:study_site}

\begin{figure}[h!]
    \centering
    \includegraphics[scale=0.7]{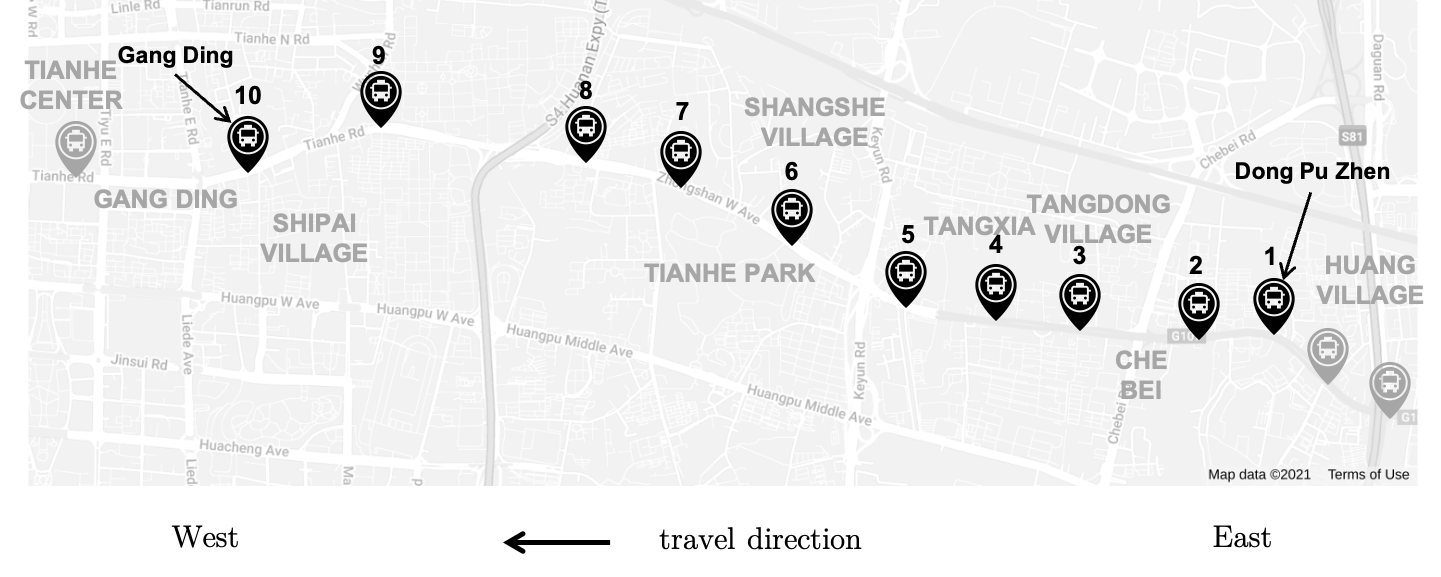}
    \caption{Layout of the 10-stop stretch of the GBRT corridor.}\label{fig:gbrt_layout}
\end{figure}

Simulations were performed for a portion of the Guangzhou Bus Rapid Transit (GBRT) corridor in China. In its entirety, the corridor consists of 26 stops spanning approximately 22.5 kms. We evaluated westbound travel (toward the city center) during the morning rush at ten consecutive, 3-berth stops shown as pins in Figure~\ref{fig:gbrt_layout}, which constitute the busiest 6.8-km stretch in the system; and did so for eight of its bus lines. Four of the lines span the entirety of the ten-stop stretch; two diverge just before reaching the western-most station labeled 10 in Figure~\ref{fig:gbrt_layout}; one spans stops 4 through 10; and the last line diverges from the corridor after visiting stop 1. Buses on the final two of these lines were the only ones not held at a simulated control point that was inserted immediately upstream of the stop labeled 1. 

\par

\subsection{Data collection and parameter estimation}\label{sec:data_collection}

Data were collected from the corridor's ten-stop stretch during the hours of 7-9 AM, and over the five weekdays spanning October 12-16, 2020. At each of the ten stops, patron boardings and alightings were counted for each bus by human observers. Boarding and alighting flows are shown in Figure~\ref{fig:stop_demand} for each stop.\footnote{Since we did not have OD data for patrons, the simulation model could not generate patron alighting flows accurately. To compensate, we factored the measured boarding flows upward by suitable amounts, to reflect the influence of alighting on bus dwell times.} Specific boarding and alighting flows for each bus line at each stop are shown in Fig.~\ref{fig:stop_demand}.
\begin{figure}[h!]
    \centering
    \includegraphics[scale=0.70]{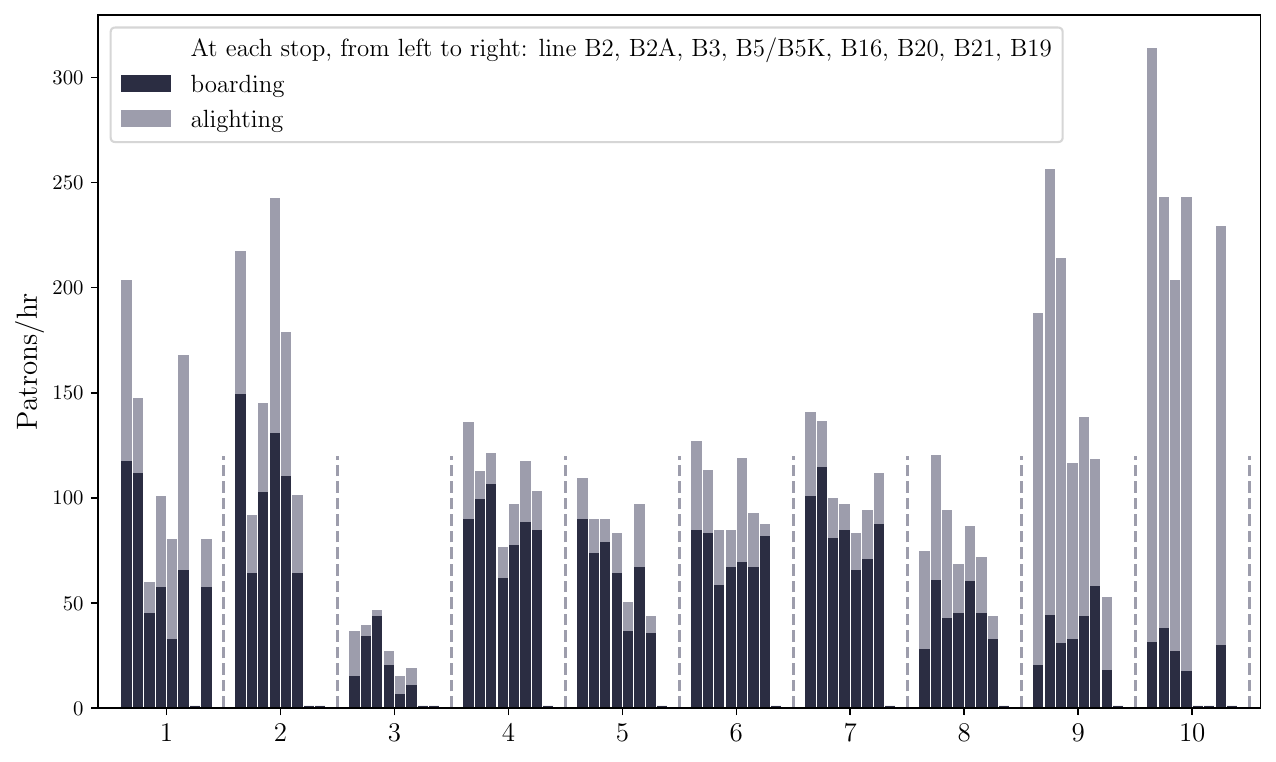}
    \caption{Boardings and alightings at each stop.}\label{fig:stop_demand}
\end{figure}\par

Our rough calculation shows that these rates are less than 2/3 of those reported prior to the COVID-19 pandemic \citep{brtdata}. We will return to this matter in later sub-sections. Still, the observed rates serve our purposes, and this too will become clear in sub-sections to follow. \par

\begin{figure}[h!]
    \centering
    \includegraphics[scale=0.62]{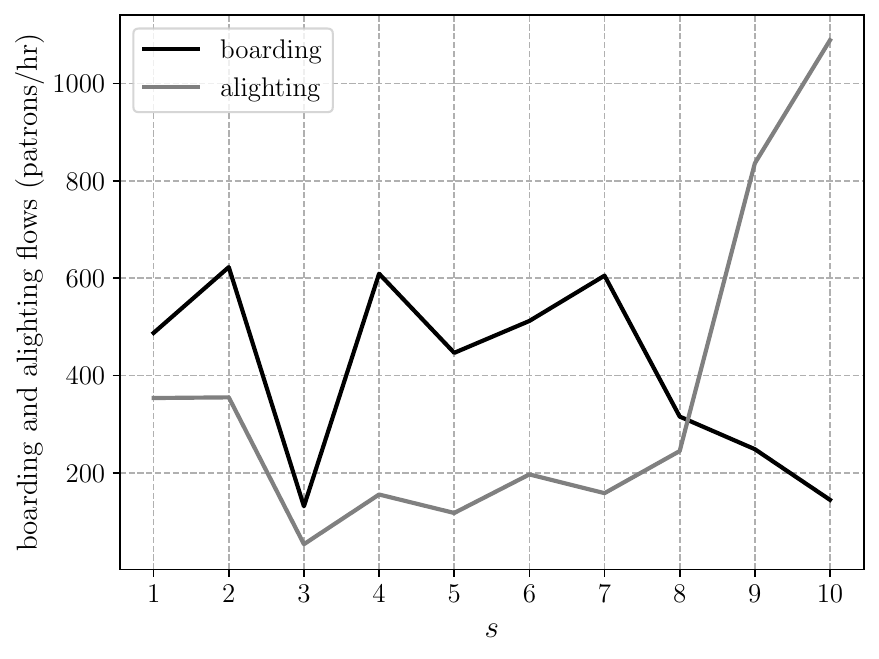}
    \caption{Boardings and alightings at each stop.}\label{fig:stop_demand}
\end{figure}\par

Bus arrival- and dwell-times at each stop were measured by the observers as well. Inter-stop travel times were collected via the buses' onboard GPS. The average headway between bus arrivals on line $l$ as they entered the corridor was used as the $H_l$. The $C_{H,l}$ was taken to be the root-mean-square deviation of the buses' arrival times on line $l$ from their scheduled times, divided by $H_l$.\par

Bus arrival- and dwell-times at each stop were measured by the observers as well. The average headway between bus arrivals on line $l$ as they enter the corridor was used as the $H_l$. The $C_{H,l}$ was taken to be the root-mean-square deviation of the buses' arrival times on line $l$ from their scheduled times, divided by $H_l$. Results are summarized in Table \ref{table:arrival_info}.\par

\begin{table}[h!]
\centering
\caption{Bus arrival time distribution parameters.}\label{table:arrival_info}
\begin{tabulary}{\textwidth}{CCCCCCCCC}
\hline
Line & B2 & B2A & B3 & B5/B5K & B16 & B20 & B21 & B19\\
$H_{l}$ (sec) & 200 & 200 & 300 & 300 & 300 & 218.2 &218.2 &480 \\
$C_{H,l}$ & 1.10 & 0.88 & 0.98 & 0.25 & 0.64 & 0.94 & 1.08 & 1.0 \\
Line group & 1 & 1 & 2 & 2 & 3 & 3 & NA & NA \\
\hline
\end{tabulary}
\end{table}\par

The unavailability of patron OD counts made it impossible to estimate the arrival rates of common-line patrons.  It was therefore assumed that the six lines spanning most or all of the ten stops are divided into $K=3$ line groups. Each group consisted of two lines that share a certain proportion of common-line patrons, $\gamma$. Hence, $\gamma=\frac{\lambda_{k,C}^{s}}{\lambda_{k,C}^{s} + \sum_{l\in\text{group}\,k}{\lambda_l^s}}$, for line group $k\in\{1,2,3\}$, and $s=1,2,\ldots,10$. We assumed $\gamma=0.5$ for our experiments, except where specified otherwise. We reckon this a conservative assumption, since the six lines share 9 or more stops.\par

Finally, the inter-stop bus travel times were collected from onboard GPS data and fitted by lognormal distributions (which exhibited higher goodness-of-fit than other distribution forms we tested). Means and standard deviations of the fitted distributions for each inter-stop link are presented in Table \ref{table:link_tt}.

%\begin{table}[h!]
%\caption{Inter-stop bus travel time distribution parameters (sec).}\label{table:link_tt}
%%\begin{tabulary}{\textwidth}{llllllllll}
%
%\begin{tabulary}{\textwidth}{lP{.062\textwidth}P{.062\textwidth}P{.062\textwidth}P{.062\textwidth}P{.062\textwidth}P{.062\textwidth}P{.062\textwidth}P{.062\textwidth}P{.062\textwidth}
%}
%\hline
%\\[-0.6em]
%Link
%& \parbox{0.05\textwidth}{\centering 1\\$\downarrow$\\2}
%& \parbox{0.05\textwidth}{\centering 2\\$\downarrow$\\3} 
%& \parbox{0.05\textwidth}{\centering 3\\$\downarrow$\\4}
%& \parbox{0.05\textwidth}{\centering 4\\$\downarrow$\\5} 
%& \parbox{0.05\textwidth}{\centering 5\\$\downarrow$\\6} 
%& \parbox{0.05\textwidth}{\centering 6\\$\downarrow$\\7} 
%& \parbox{0.05\textwidth}{\centering 7\\$\downarrow$\\8} 
%& \parbox{0.05\textwidth}{\centering 8\\$\downarrow$\\9} 
%& \parbox{0.05\textwidth}{\centering 9\\$\downarrow$\\10} \\
%\\[-0.6em]
%\hline
%%\textbf{Input parameters}& \\
%%\multicolumn{8}{c}{}\\[0.15em]
%Mean & 53.1 & 58.1 & 24.2 & 32.5 & 102.3 & 35.5 & 69.6 & 90.6 & 87.5 \\
%Standard deviation & 11.3 & 22.5 & 9.5 & 8.5 & 34.7 & 8.5 & 24.0 & 25.5 & 41.5 \\
%\hline
%\end{tabulary}
%\end{table}

\begin{table}[h!]
\caption{Inter-stop bus travel time distribution parameters (sec).}\label{table:link_tt}
\begin{tabulary}{\textwidth}{llllllllll}
\hline
\\[-0.6em]
Link
& \parbox{0.05\textwidth}{\centering 1\\$\downarrow$\\2}
& \parbox{0.05\textwidth}{\centering 2\\$\downarrow$\\3} 
& \parbox{0.05\textwidth}{\centering 3\\$\downarrow$\\4}
& \parbox{0.05\textwidth}{\centering 4\\$\downarrow$\\5} 
& \parbox{0.05\textwidth}{\centering 5\\$\downarrow$\\6} 
& \parbox{0.05\textwidth}{\centering 6\\$\downarrow$\\7} 
& \parbox{0.05\textwidth}{\centering 7\\$\downarrow$\\8} 
& \parbox{0.05\textwidth}{\centering 8\\$\downarrow$\\9} 
& \parbox{0.05\textwidth}{\centering 9\\$\downarrow$\\10} \\
\\[-0.6em]
\hline
%\textbf{Input parameters}& \\
%\multicolumn{8}{c}{}\\[0.15em]
Mean & 53.1 & 58.1 & 24.2 & 32.5 & 102.3 & 35.5 & 69.6 & 90.6 & 87.5 \\
Standard deviation & 11.3 & 22.5 & 9.5 & 8.5 & 34.7 & 8.5 & 24.0 & 25.5 & 41.5 \\
\hline
\end{tabulary}
\end{table}

\section{Numerical Analysis}\label{sec:numerical_analysis}

Simulation outcomes in Section \ref{sec:vicious_cycle} confirm that the vicious cycle described in Section \ref{sec:introduction} can occur, and that its effects can be abated via bus holding. Outcomes in Section \ref{sec:benefits_long_busy} reveal how holding can reduce bus delays corridor-wide in corridors with sufficient numbers of tandem stops and sufficiently high patron demands. Outcomes in Section \ref{sec:shorten_hold_time} reveal the substantial advantages of reducing the minimum headway threshold for bus holding. Finally, Section~\ref{sec:comparison} shows the advantage of our modified holding strategy compared against those in the literature.

%Outcomes in Section \ref{sec:common_line} show how holding buses by group makes sense when a corridor serves common-line patrons in sufficient number. 

\subsection{Vicious cycle unveiled and abated}\label{sec:vicious_cycle}

\begin{figure}[h!]
\centering
\begin{subfigure}[t]{0.45\columnwidth}
        \includegraphics[width=\columnwidth]{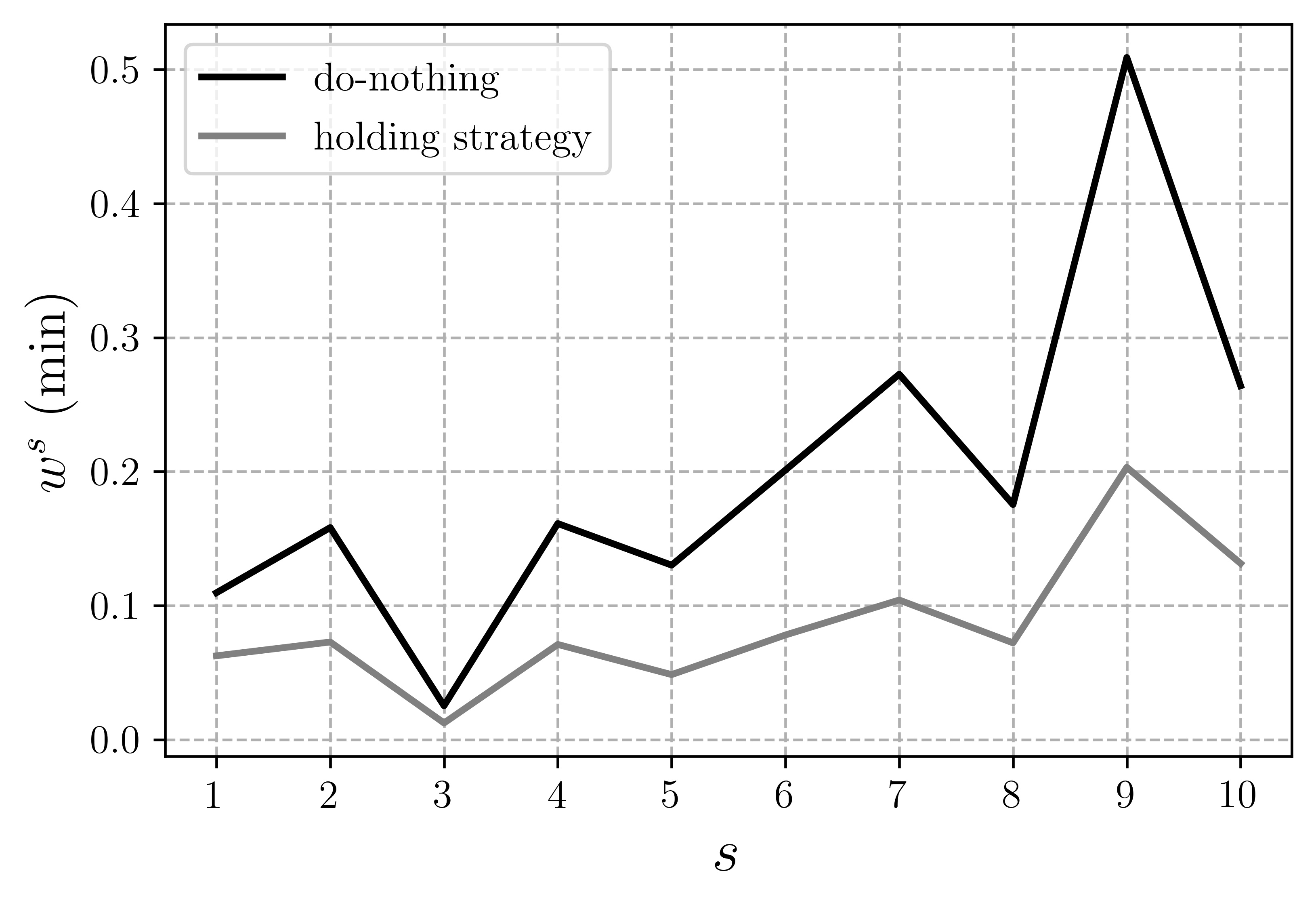}
        \caption{\centering bus delays under present-day patron demands}
        \label{fig:w_100}
    \end{subfigure}
	\begin{subfigure}[t]{0.45\columnwidth}
        \includegraphics[width=\columnwidth]{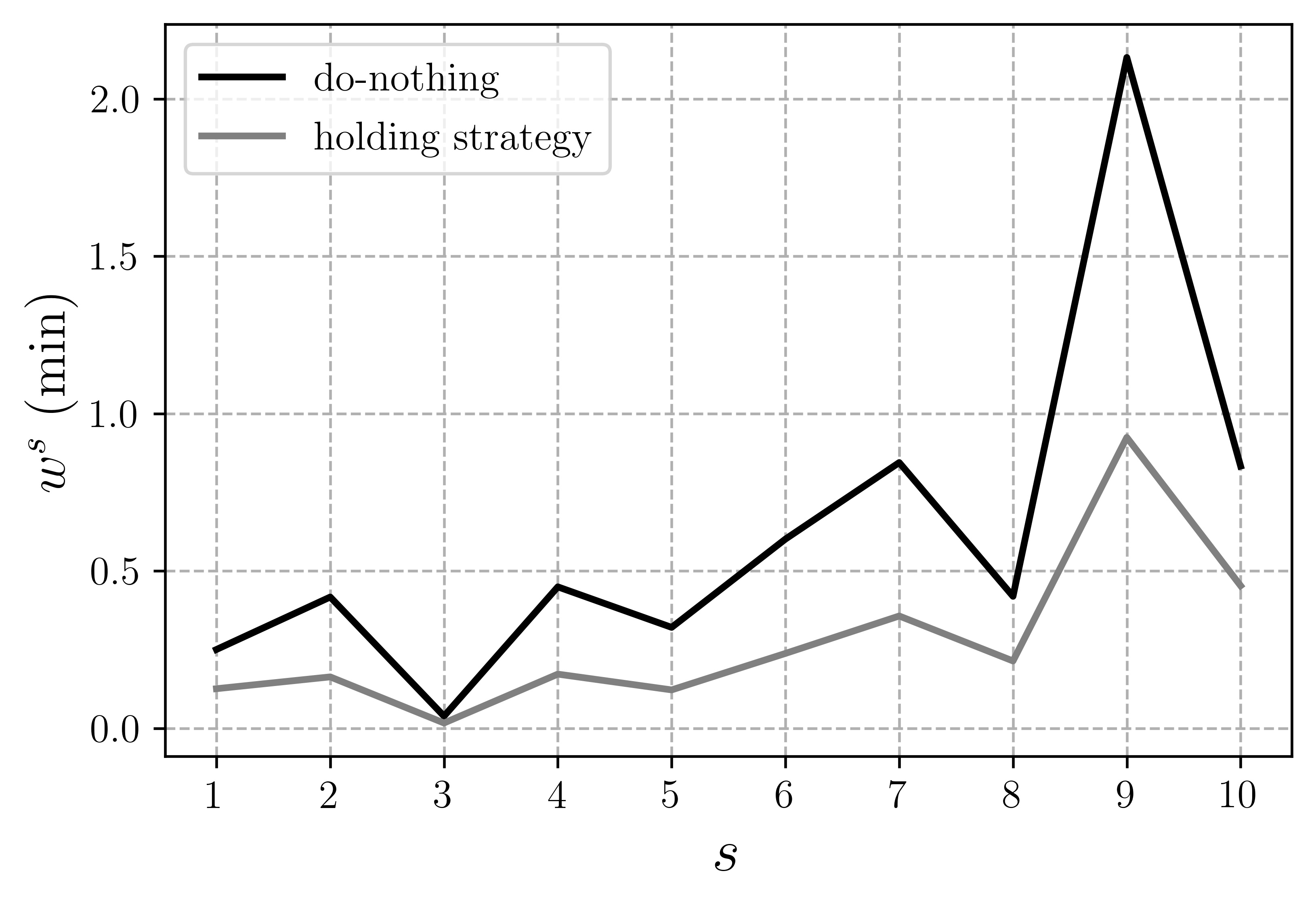}
        \caption{\centering bus delays under pre-pandemic demands}
        \label{fig:w_150}
    \end{subfigure}
    \caption{$w^s$ with and without holding ($\eta=0.9$).}
\end{figure}

The curves in Figure~\ref{fig:w_100} show average bus delay at each stop, $w^s$, for the patron demands revealed via our boarding and alighting counts. The bold, upper curve displays simulated outcomes in the absence of holding. It shows that abrupt drops in $w^s$ occur at stops 3, 8, and 10. Those stops are the ones characterized by lower boardings and alightings, as can be confirmed from Figure~\ref{fig:stop_demand}. Despite this occasional relief, the bold curve in Figure~\ref{fig:w_100} trends upward: note, for example, the larger $w^s$ at downstream-most stop 10 compared against those at stops 1 and 2 upstream.\footnote{Queueing delay is positively correlated with dimensionless ``traffic intensity'', defined in our case as the product of bus flow and average dwell time \citep{almeida2018note,shen2023efficient}. However, the traffic intensity of stop 10 (0.71) is substantially lower than those of stops 1 and 2 (0.78 and 0.81, respectively). This indicates that the larger delay at stop 10 is due to the highly-varied headways, and not to excessive boardings or alightings there.} Also note that patron boarding and alighting flows remained fairly constant across stops 4-7, as evident in Figure~\ref{fig:stop_demand}. Yet, the bold curve in Figure~\ref{fig:w_100} trends upward as buses move forward, traversing these stops.\par

These trends unveil the vicious cycle previously described. Not surprisingly, simulations also reveal that variations in bus arrival headways at stops consistently increase with $s$.\footnote{Simulation tests were also conducted for homogeneous corridors where each stop had the same patron demand. Results under these idealized conditions demonstrate more emphatically that both bus delay and headway variation increase monotonically with $s$.} \par

\par

The lighter-drawn, lower curve in Figure~\ref{fig:w_100} presents the $w^s$ from simulations in which the proposed holding strategy was put in place, and buses in each of six lines were held at the control point until a minimum headway of $\eta\cdot H_l$ elapsed between consecutive departures from the same line, where $\eta$ is the holding time adjustment factor described in Section \ref{sec:bus_dynamics}. (More will be said momentarily as regards selection of $\eta$, which was 0.9 in the present case.) Visual inspection of both curves in the figure confirms that holding can substantially retard the vicious cycle by regularizing bus headways at the corridor's entry. (Headway variations in all the downstream stops were diminished as well.) Note how greater delay savings are achieved at downstream stops, where bus queueing is more severe and the reduced headway variations thus bring more benefits.\par

The curves in Figure~\ref{fig:w_150} were constructed by increasing observed patron demand by 50\%, to bring demand closer to the rate reported for the site prior to the pandemic. Visual inspection of Figures \ref{fig:w_100} and \subref{fig:w_150} reveals how both the vicious cycle, and its retardation through holding grow with demand for bus travel.\par

\subsection{Benefits for long, busy corridors}\label{sec:benefits_long_busy}

Findings of the last section indicate that bus holding might save the bus delay accumulated over the entire corridor,\footnote{Even if only part of the holding delay is offset by the delay savings at downstream stops, holding is still beneficial due to its headway-regularizing effect.} where corridors are long and hold sufficient stops. We examine this matter using cumulative bus delay, $W^s$, as the metric.\par

\begin{figure}[h!]
\centering
\begin{subfigure}[t]{0.45\columnwidth}
        \includegraphics[width=\columnwidth]{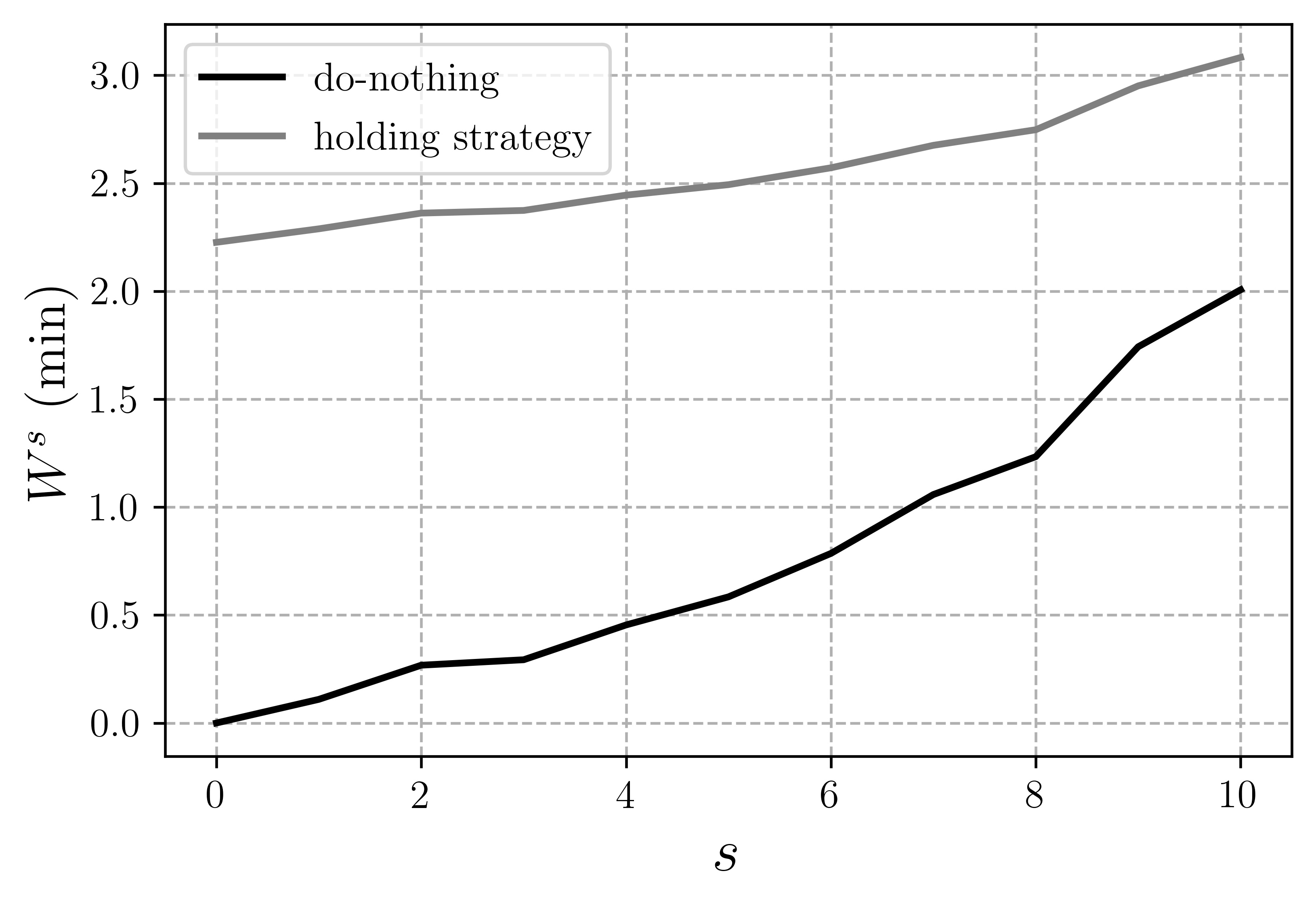}
        \caption{present-day patron demands}
        \label{fig:cw_100}
    \end{subfigure}
	\begin{subfigure}[t]{0.45\columnwidth}
        \includegraphics[width=\columnwidth]{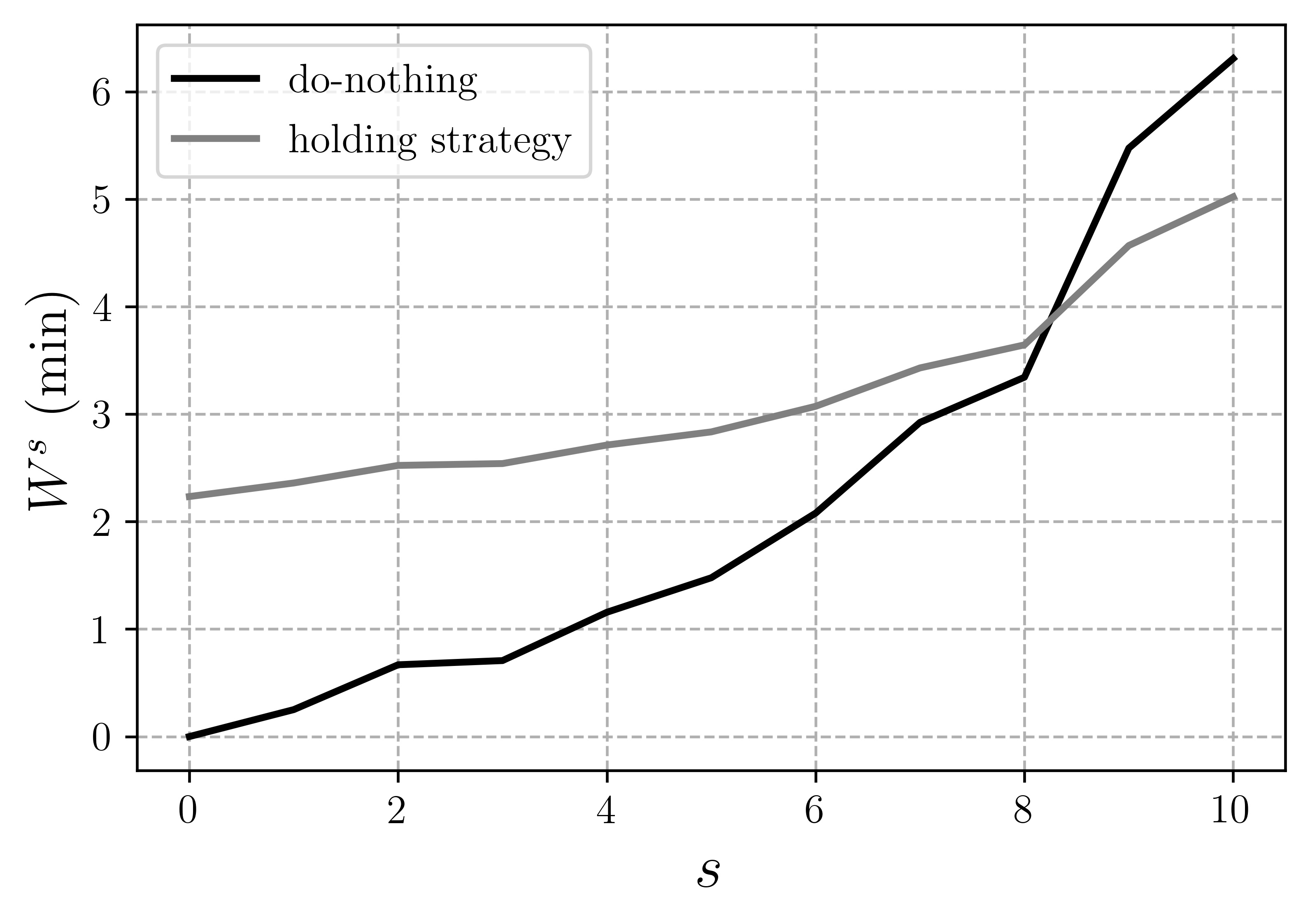}
        \caption{pre-pandemic demands}
        \label{fig:cw_150}
    \end{subfigure}
    \caption{$W^s$ with and without holding ($\eta=0.9$).}
\end{figure}

Curves for present-day patron demands are shown in Figure~\ref{fig:cw_100}. Holding delay is the intercept at $s=0$ for the lighter-drawn curve (approximately 2.2 mins/bus). Note how that curve consistently lies above its bold counterpart despite the diminishing vertical displacements between the two curves; i.e., the holding strategy results in larger $W^s$ for all $s\leq 10$. The light curve might have fallen below the bold one, but only for long corridors with $s>10$.\par

\par

The curves in Figure~\ref{fig:cw_150} tell a different story, however. They were constructed from simulations in which patron demands were increased by 50\% to approach pre-pandemic levels. Note how holding now produces net delay savings for corridors with $s>8$.\par 

Not surprisingly, simulated outcomes under yet higher patron demands (not shown for brevity) revealed that holding can save bus delays on even shorter corridors. Hence we see the twofold value of holding when patron demands are sufficiently large and corridors contain sufficient numbers of stops: the strategy can reduce not only headway variations, but bus delays as well.

\subsection{Shortening the headway threshold for holding}\label{sec:shorten_hold_time}

We next demonstrate benefits of diminishing bus holding times by using an $\eta<1$. We do so by once again increasing observed patron demands by 50\% to bring them to pre-pandemic levels.\par

\begin{figure}[h!]
    \centering
    \includegraphics[scale=0.65]{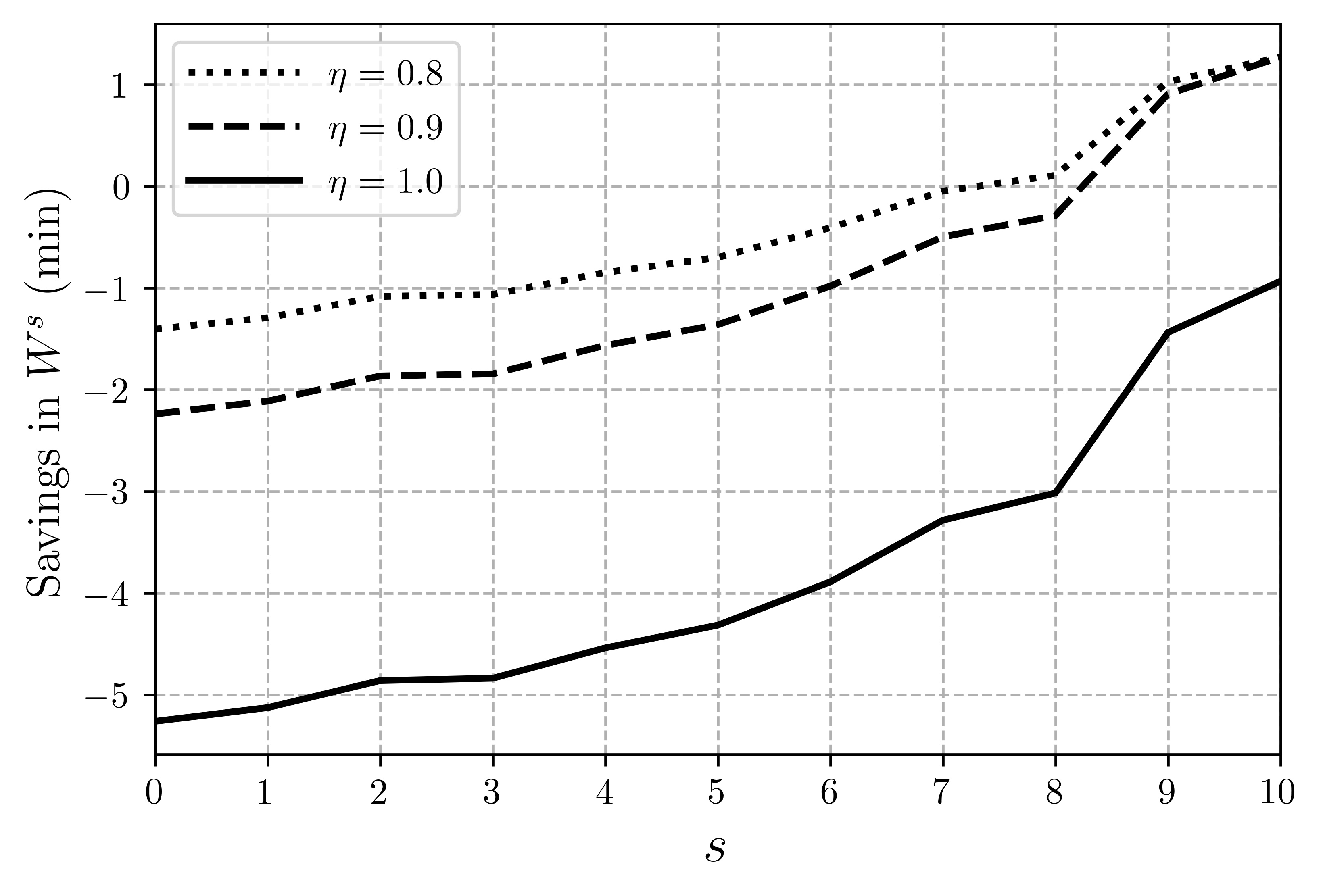}
    \caption{Effect of $\eta$.}\label{fig:sa_hold_eta}
\end{figure}

The lowest, solid curve in Figure~\ref{fig:sa_hold_eta} displays the savings in $W^s$ (measured relative to the do-nothing case) when $\eta=1$. Note how holding in this case is inferior to the do-nothing option, at least for the ten-stop corridor stretch examined here.\par

In contrast, the middle, dashed curve in Figure~\ref{fig:sa_hold_eta} shows that considerable benefits occur by reducing the minimum headway threshold, such that $\eta=0.9$. Note from the curve's intercept how a 10\% reduction in $\eta$ produces a 58\% reduction in holding time.
\par

Benefits dwindle as $\eta$ takes smaller values, as evident from the figure's upper-most, dotted curve corresponding to $\eta=0.8$. Despite a moderate additional reduction in the holding delay, savings in bus queueing delays at downstream stops also decrease. The latter occurs because holding becomes less effective in regularizing headways as $\eta$ diminishes.\par

\subsection{Comparisons with other holding strategies}\label{sec:comparison}
We compare our simple, modified strategy against six previously-formulated ones, including the conventional schedule-based holding strategy in \citet{boyle2009controlling}, and those proposed in \citet{daganzo2009headway}, \citet{xuan2011dynamic}, \citet{daganzo2011reducing}\footnote{\citet{daganzo2011reducing} introduced a two-way-looking speed control strategy that relies on bus spacings instead of headways. For simplicity, here we employ the Eulerian version of this strategy, with holding as the control method in place of speed adjustments. The same modification was also adopted by the comparative analyses in \citet{xuan2011dynamic} and \citet{berrebi2018comparing}.}, \citet{bartholdi2012self}, and \citet{berrebi2015real}. All of the six existing strategies and our own are applied to the GBRT corridor under pre-pandemic demands. A single control point is used (like in \citealp{berrebi2018comparing}). For strategies in \citet{daganzo2009headway}, \citet{xuan2011dynamic}, and \citet{daganzo2011reducing}, this control point is located at the corridor's first stop since these strategies require the use of the demand rates at the stop where holding is applied. For the other strategies (including ours), the control point is located upstream of the corridor's first stop. Similar to \citet{berrebi2018comparing}, slack is ignored in our comparisons. Holding time formulas for the $j$-th bus on line $l$ under the six existing strategies are presented in Table \ref{table:holding_methods}. When these formulas yielded non-positive values, holding was not applied.\par
\begin{table}[h!]
\caption{Holding methods}\label{table:holding_methods}
\footnotesize
    \begin{tabular}{cc}
        \hline
        Holding method & Holding time \\ \hline
Conventional schedule-based method \citep{boyle2009controlling} & $\bar{a}_{l,j}^0 - a_{l,j}^0$ \\
\citet{daganzo2009headway} & $(\alpha+\hat{\beta}_{l}^{s=1}) \left(H_l-( d_{l,j}^{s=1}-d_{l,j-1}^{s=1} )\right)$ \\
The simple control law in \citet{xuan2011dynamic} &$\hat{\beta}_{l}^{s=1}\left(H_l-(d_{l,j}^{s=1}-d_{l,j-1}^{s=1})\right) + \alpha\left(\bar{d}_{l,j}^{s=1}-d_{l,j}^{s=1}\right)$ \\
\citet{daganzo2011reducing} & $(\alpha+\hat{\beta}_{l}^{s=1}) \left(H_l-( d_{l,j}^{s=1}-d_{l,j-1}^{s=1} )\right) - \alpha (H-(\hat{d}_{l,j+1}^{s=1}-d_{l,j}^{s=1}))$ \\
\citet{bartholdi2012self} & $\max\left\{H_l-(a_{l,j}^{0}-a_{l,j-1}^{0}), \alpha\left( \hat{a}_{l,j+1}^{0}-a_{l,j}^{0} \right) \right\}$ \\
\citet{berrebi2015real} & $\frac{\max_{r\in[1,\ldots,M]}\left\{ \frac{\hat{a}_{l,j+r}^{0}- a_{l,j}^0}{r}\right\} - \left(a_{l,j}^{0}-a_{l,j-1}^{0}\right)}{1+\left(\argmax_{r\in[1,\ldots,M]}\left\{ \frac{\hat{a}_{l,j+r}^{0}- a_{l,j}^0}{r}\right\}\right)^{-1}}$\\
        \hline
    \end{tabular}
\end{table}

\normalsize
The notations used in Table \ref{table:holding_methods} are defined as follows: $\bar{a}_{l,j}^0$ is the scheduled arrival time of bus $j$ on line $l$ at the control point; $\hat{\beta}_{l}^{s=1}$ is the product of the estimated patron arrival rate and boarding time per patron at the first stop on line $l$; $\alpha$ is the control parameter; $\bar{d}_{l,j}^{s=1}$ is the scheduled departure time of bus $j$ from the first stop on line $l$; $\hat{d}_{l,j+1}^{s=1}$ is the predicted departure time of the next bus from the first stop on line $l$; $\hat{a}_{l,j+r}^{0}$ is the predicted arrival time of the next $r$-th bus at the control point on line $l$; and $M$ is the number of buses to follow, whose arrival times will be predicted in ways to be described.\par

\begin{figure}[h!]
    \centering
    \includegraphics[scale=0.56]{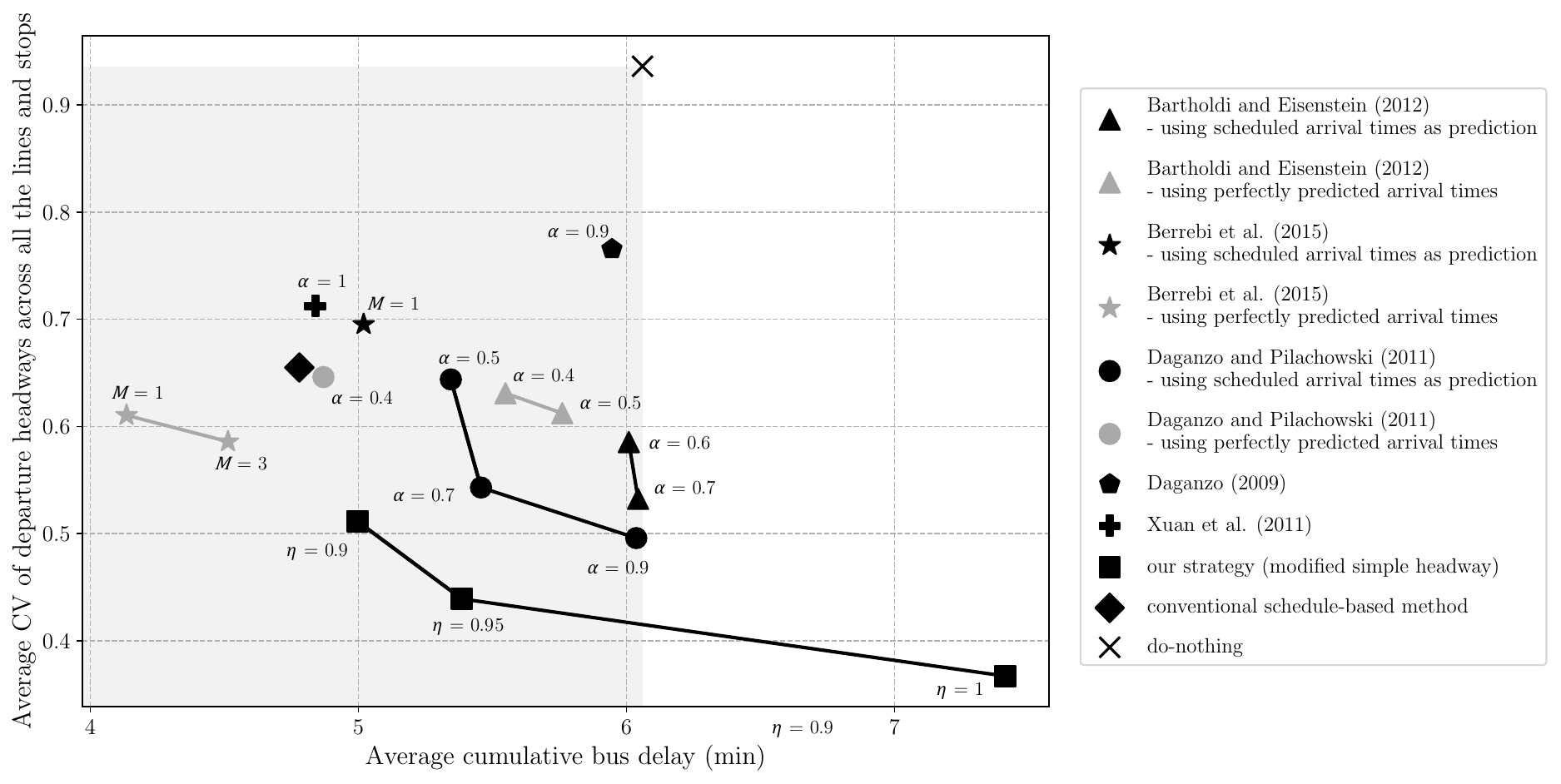}
    \caption{\centering Performance of different holding strategies in terms of headway variations and cumulative bus delays.}\label{fig:comparison}
\end{figure}

Figure \ref{fig:comparison} plots the coefficient of variation (CV) in departure headways on its y-axis. These are averaged across all the lines and stops. The average cumulative bus delay in the corridor is shown on the figure's x-axis. The data are shown for all the strategies, along with the attendant control parameters. The two metrics are the most commonly used in the literature (e.g., \citealp{estrada2016bus} and \citealp{berrebi2018comparing}). For strategies in \citet{daganzo2009headway}, \citet{xuan2011dynamic}, and \citet{daganzo2011reducing}, the estimated demand parameter, $\hat{\beta}_l^s$, is assumed to equal the actual demand. Strategies in \citet{daganzo2011reducing}, \citet{bartholdi2012self}, and \citet{berrebi2015real} rely on predicted arrival times of following buses at the control point. Two kinds of predictions are used: (i) perfect prediction, assuming that buses' future arrival times can be perfectly forecasted; and (ii) bus arrival times are predicted to follow their schedules. Note that (i) represents a best-case scenario while (ii) a worst-case scenario since scheduled arrival times reflect the mean values.

%Two kinds of predictions are used: (i) perfect prediction, assuming that buses' future arrival times can be perfectly forecasted, perhaps thanks to technologies; and (ii) bus arrival times are predicted to follow their schedules.\par

%For strategies in \citet{daganzo2009headway}, \citet{xuan2011dynamic} and \citet{daganzo2011reducing},
%in \citet{daganzo2011reducing}, \citet{bartholdi2012self} and \citet{berrebi2015real} 

First note that most points in Figure~\ref{fig:comparison} lie in the lightly-shaded region, below and to the left of the do-nothing case, shown with an X. This means that all the said holding strategies can retard the vicious cycle and save bus delays in a congested corridor.\par
\par

Further inspection of Figure~\ref{fig:comparison} shows that our simple modified strategy produces the lowest variation in bus headway. Our method also performs well in terms of reducing bus delay. Note in this context the value of choosing $\eta < 1$.\par

There are other strategies that can do as well, or better, in this regard, however. Included in this list are the conventional schedule-based method and \citet{xuan2011dynamic}. The strategies in \citet{daganzo2011reducing} and \citet{berrebi2015real} can also save more delay than can our strategy, but only when they benefit from perfect predictions of bus arrival times. Even with advanced technology, perfect prediction can be elusive when corridors congest and impede bus movements. In contrast, recall that our simple strategy requires only real-time measurements of bus arrival times.

\section{Common-line patrons and holding by group}\label{sec:common_line}
Common-line patrons are known to trigger greater variations in bus headway and exacerbate bunching \citep{schmocker2016bus}. To mitigate this adverse effect, buses can be held and released by line group instead of by individual line. Thus, we propose the following modification of the simple headway-based strategy: Arriving buses that belong to the same line group $k$ will queue in a same lane at the control point (see Figure~\ref{fig:sorting_point_by_line}), and be released with a minimum headway of $\sfrac{1}{\sum_{l\in \text{group}\,k}{f_l}}$, i.e., the joint headway in line group $k$.\par
\begin{figure}[h!]
    \centering
    \includegraphics[scale=0.66]{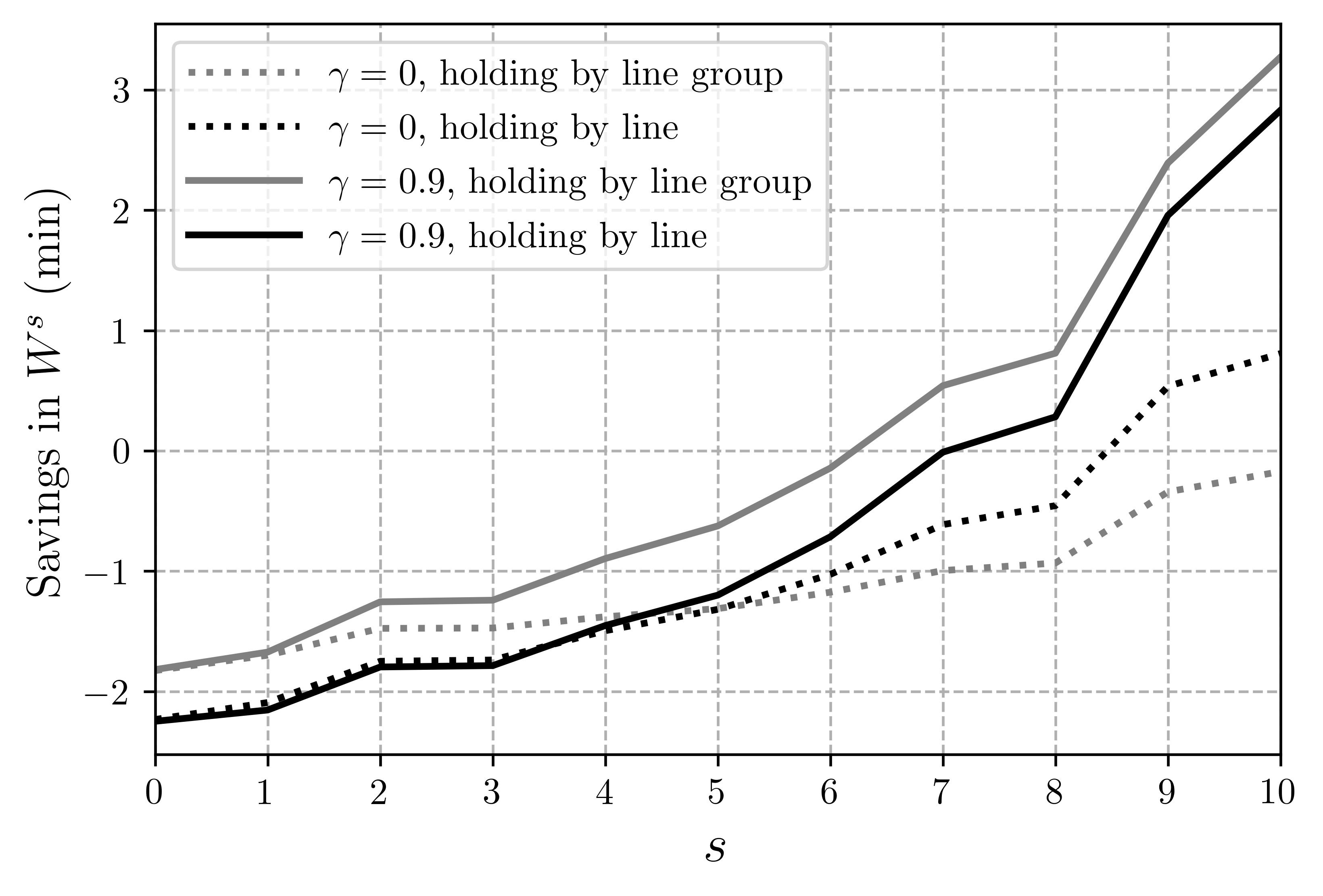}
    \caption{Effect of $\gamma$ on holding strategy.}\label{fig:sa_hold_gamma}
\end{figure}

 
The above modification can produce favorable results when corridors serve common-line patrons in high numbers. To illustrate, the two solid curves in Figure~\ref{fig:sa_hold_gamma} display the $W^s$ that would be saved through holding if the corridor had a large percentage of these patrons, such that $\gamma=0.9$. The lighter-drawn of these curves plots outcomes when buses are held by group. Its bolder counterpart displays what happens when holding occurs by line.\par

%This modification can reduce holding delay because the headways exhibited by consecutively-arriving buses across multiple lines of the same group are smaller than those from a single line within the group. 
%This modification can further reduce bus delays experienced along a corridor. It can also reduce holding delay because the headways exhibited by consecutively-arriving buses across multiple lines of the same group are smaller than those from a single line within the group.\par

\par

Note that for the large $\gamma$, either form of holding can save bus delays. This is not surprising: with large $\gamma$, some buses are likely to serve greater numbers of patrons, leading to more varied dwell times across buses. This, in turn, further triggers larger bus delays \citep{gu2011capacity,gu2015models} and more bunching \citep{schmocker2016bus}. Holding is thus especially valuable.\par
 
The solid curves in Figure~\ref{fig:sa_hold_gamma} also make clear that holding buses by line group is the preferred option. This is because holding by group produces lower holding delays (thanks to the smaller bus arrival headways within each line group), and with large $\gamma$, the regularized bus headways within each line group also engender more delay savings at stops 1-6. On the downside, holding buses by group produces less delay savings at downstream stops 7-10. This is due to the larger line-specific headway variations created by that strategy.\footnote{One may be concerned that holding by line group might increase the variations in line-specific headways and thus exacerbate bunching once buses are traveling on line portions that have diverged from the corridor. However, our simulations confirm that holding by line group still reduces headway variation over the entirety of each line, although not as much as holding by line alone.}\par

Things are different if the corridor serves low numbers of common-line patrons. To see this, the dotted curves in Figure~\ref{fig:sa_hold_gamma} display the $W^s$ saved if the corridor had no common-line patrons, such that the ratio of common-line patrons shared by bus lines in a line group, $\gamma=0$. The lighter of the dotted curves shows that holding buses by group in this case is worse than doing nothing. This is because holding by group would now be inefficient in regularizing headways.\par

Holding by line alone offers some improvement in this case, as evident from the figure's boldly-drawn, dotted curve. Yet, the benefits of holding clearly degrade when common-line patrons are few.

\section{Conclusions}\label{sec:conclusion}
Physically-realistic models were formulated and used to simulate bus movements along part of the BRT corridor in Guangzhou, China. Five findings of interest were unveiled. First, the simulations confirmed that queueing at curbside stops can trigger a vicious cycle, whereby buses encounter worsening conditions as they proceed along a corridor. Second, this cycle can be mitigated by temporarily holding buses at the corridor's upstream end. Third, savings in bus delay can outweigh the delay due to holding when long corridors contain a sufficient number of stops. For example, by using our simple modified method of holding according to headway, more than 20\% of the overall bus delay can be saved on our Guangzhou site. Even greater savings are expected for busier corridors, such as the Santo Amaro-Nove de Julho-Centro corridor mentioned in the introduction, which accommodates a significantly higher patron flow compared to the studied GBRT corridor. Note that these three findings are unique to congested bus corridors, since holding will only increase bus delays in corridors that are free of bus queues. These findings also indicate that bus commercial speed would improve under our strategy, so that fleet size can be reduced. In addition, holding can reduce patrons' delays in two ways: (i) by reducing in-vehicle delays as bus delays diminish; and (ii) by reducing waiting times as bus headways become less varied \citep{newell1964maintaining}. Fourth, our simple strategy outperforms other strategies in the literature in regularizing headways, and thus in reducing bus delays in a congested corridor. Our strategy becomes the best in this regard when future bus arrival times cannot be accurately predicted. Fifth and last, simulations also showed that grouping buses and holding them by group can be effective when large proportions of patrons can choose buses from among multiple lines.\par

Though the present case study features only a single control point, our models can be readily extended to accommodate multiple such points that might be suitably placed along longer and busier corridors. The best control point locations for a given corridor can be determined by comparing the simulation outcomes under different location plans. Simulation can also be used to determine the best grouping plan of bus lines when buses are held by line group.\par

Further note (in Section \ref{sec:comparison}) that more sophisticated adaptive holding strategies often performed worse than did both the conventional schedule-based method \citep{boyle2009controlling} and our modification to the simple headway-based holding method. This outcome suggests need for further research, to determine if novel, more effective control strategies might be developed by leveraging advanced technologies. These new strategies might adaptively control bus speeds in ways akin to \citet{he2019approach} or prioritize buses at signalized intersections as described in \citet{estrada2016bus} and \citet{anderson2020effect}. Our models presented in this paper could be used to evaluate these new control schemes.

%This calls for developing novel and more effective control strategies leveraging emerging technologies, which are suitable for congested corridors. Our models can be used to study those novel forms of strategies, including adaptive bus speed control \citep{he2019approach} and novel signal priority schemes \citep{estrada2016bus, anderson2020effect}.\par

\section*{Acknowledgements}
This work was supported by the National Natural Science Foundation of China (Project No. 72201214), the General Research Funds (No.\,15217415, 15224317) provided by the Research Grants Council of Hong Kong, the Sichuan Science and Technology Program (Project No. 2023NSFSC1035) and the Fundamental Research Funds for the Central Universities under Grant JBK23YJ01. The authors thank the nine students from South China University of Technology who helped collect data at the Guangzhou BRT and extracted bus trajectories from GPS data.

%General Research Funds (No.\,15217415, 15224317) provided by the Research Grants Council of Hong Kong. 

\newpage
\begin{appendices}
\setcounter{table}{0} \renewcommand{\thetable}{A.\arabic{table}}
\section{Table of Notations}\label{apdx:notation}
\begin{table}[h!]
\centering
\caption{List of notations}\label{table:notations}
\begin{tabulary}{\textwidth}{p{.15\textwidth}p{.85\textwidth}}
\hline
Notation & Description \\ \hline
%\multicolumn{2}{l}{\textbf{Variables}}\\[0.15em]
%$b^s$ & Number of patrons onboard when arriving at stop $s$\\[0.07em]
$a_{l,j}^0$ & Arrival time of the $j$-th bus on line $l$ to the corridor's upstream end \\[0.07em]
$\bar{a}_{l,j}^0$ & Scheduled arrival time of the $j$-th bus on line $l$ to the corridor's upstream end \\[0.07em]
$\hat{a}_{l,j+r}^{0}$ & Predicted arrival time of the next $r$-th bus to the corridor's upstream end when the $j$-th bus on line $l$ arrives \\[0.07em]
$a_{l,j}^s$ & Arrival time of the $j$-th bus on line $l$ to stop $s$ \\[0.07em]
$\alpha$ & Control parameter in \citet{daganzo2009headway}, \citet{xuan2011dynamic}, \citet{daganzo2011reducing} and \citet{bartholdi2012self}. \\[0.07em]
$\beta^s_l$ & A dimensionless parameter accounting for the marginal bus dwell time resulting from a unit increase in headway on line $l$ at stop $s$\\[0.07em]
$b_{l,j}^s$ & In-berth delay of the $j$-th bus on line $l$ at stop $s$ \\[0.07em]
$c^s$ & Number of berths at stop $s$\\[0.07em]
%$C$ & Bus capacity\\[0.07em]
$C_{H,l}$ & Coefficient of bus arrival time deviation at the corridor's upstream end for line $l$.\\[0.07em]
%${C}_{A}^{s}$ & Coefficient of variation in the buses' arrival headways at stop $s$, averaged across all lines\\[0.07em]
%${C}_{D}^{s}$ & Coefficient of variation in the buses' departure headways at stop $s$, averaged across all lines\\[0.07em]
%${C}_{E}^{s}$ & Coefficient of variation in the buses' entry headways at stop $s$, averaged across all lines\\[0.07em]
$\delta_a^s$ & Boarding time per patron at stop $s$\\[0.07em]
$\delta_b^s$ & Alighting time per patron at stop $s$\\[0.07em]
$d_{l,j}^s$ & Departure time of the $j$-th bus on line $l$ from stop $s$\\[0.07em]
$\bar{d}_{l,j}^s$ & Scheduled departure time of the $j$-th bus on line $l$ from stop $s$\\[0.07em]
$\hat{d}_{l,j+1}^{s}$ & Predicted departure time of the next bus from stop $s$ when the $j$-th bus on line $l$ departs\\[0.07em]
$\eta$ & Holding time adjustment factor\\[0.07em]
$f_l$ & Bus flow of line $l$\\[0.07em]
$\gamma$ & Ratio of common-line patrons shared by bus lines in a line group\\[0.07em]
$h_{l,j}^s$ & Number of alighting patrons for the $j$-th bus on line $l$ at stop $s$ \\[0.07em]
$H_l$ & Scheduled headway of line $l$\\[0.07em]
$\lambda_{k,C}^s$ & Common-line patron arrival rate for line group $k$ at stop $s$\\[0.07em]
$\lambda_{l}^s$ & None-common-line patron arrival rate for line $l$ at stop $s$\\[0.07em]
$L$ & Number of bus lines\\[0.07em]
$\mu_{T}^s$ & Mean bus travel time from stop $s$ to stop $s+1$ \\[0.07em]
$M$ & Number of following buses to be predicted in \citet{berrebi2015real}. \\[0.07em]
$K$ & Number of line groups\\[0.07em]
$N$ & Number of stops in the corridor\\[0.07em]
$p_{l,j}^s$ & Number of boarding patrons for the $j$-th bus on line $l$ at stop $s$ \\[0.07em]
$q_{l,j}^s$ & Queueing delay of the $j$-th bus on line $l$ at stop $s$\\[0.07em]
$\sigma_{T}^s$ & Standard deviation of bus travel time from stop $s$ to stop $s+1$ \\[0.07em]
$S_{l,j}^s$ & Dwell time of the $j$-th bus on line $l$ at stop $s$ \\[0.07em] 
%$\theta$ & Amplifying factor of patron demand in the real-world case study\\[0.07em]
$\tau$ & Time lost due to bus deceleration and acceleration, and door opening and closing at a stop\\[0.07em]
$t_{l,j}^s$ & Inter-stop travel time of the $j$-th bus on line $l$ from stop $s$ to stop $s+1$\\[0.07em]
$w_{l,j}^0$ & Holding delay of the $j$-th bus on line $l$ at the control point \\[0.07em]
$w^s$ & Average bus delay at stop $s$ \\[0.07em]
$W^s$ & Average cumulative delay per bus from the corridor's upstream end to stop $s$ \\[0.07em]

\hline
\end{tabulary}
\end{table}

\section{Derivation of Equation \eqref{eq:approx_delay}}\label{apdx:approx_delay}
\setcounter{equation}{0} \renewcommand{\theequation}{B.\arabic{equation}}
\setcounter{figure}{0} \renewcommand{\thefigure}{B.\arabic{figure}}

To derive the approximation, assume first that buses on line $l\in\{1,2,\ldots,L\}$ are released from the control point in the same order as they are scheduled; i.e., bus $j+1$ is always released later than bus $j$, even if the Gaussian arrival times, are generated such that $a_{l,j}^0 > a_{l,j+1}^0$. This approximation is conservative since it slightly exaggerates holding delays.\par

Under this assumption, the release time of bus $j$ from the control point, $a_{l,j}^1$, is given by:\par
\begin{align*}
	a_{l,j}^1 &= \text{max}\{a_{l,j}^0, a_{l,j-1}^1+H_l\} = \text{max}\{a_{l,j}^0, \text{max}\{a_{l,j-1}^0,a_{l,j-2}^1+H_l\} + H_l\} \\
	&= \text{max}\{a_{l,j}^0, a_{l,j-1}^0+H_l,a_{l,j-2}^1+2H_l\} = \ldots \\
	&= \text{max}\{a_{l,j}^0, a_{l,j-1}^0+H_l,a_{l,j-2}^0+2H_l,\ldots, a_{l,1}^0+(j-1)H_l\}.
\end{align*}
Note that $a_{l,j}^0, a_{l,j-1}^0+H_l,a_{l,j-2}^0+2H_l,\ldots,a_{l,1}^0+(j-1)H_l$ are $j$ independent and identically distributed (i.i.d.) random variables following a Gaussian distribution, with mean $jH_l$ and standard deviation $C_{H,l}H_l$. Thus, $a_{l,j}^1$ is the last (largest) order statistic of $j$ i.i.d. Gaussian random variables. A very good approximation to the mean of this order statistic, as per \citet{elfving1947asymptotical}, is:
\begin{equation}
	E[a_{l,j}^1|j] \approx jH_l+C_{H,l}H_l\Phi^{-1} \left( \frac{j-\frac{\pi}{8}}{j-\frac{\pi}{4}+1} \right),
\end{equation}
where $\Phi^{-1}(\cdot)$ is the inverse function of the standard normal CDF.\par
Hence, from (\ref{eq:first_stop_arrive_time}) we have:

\begin{equation}
	E[w_{l,j}^0|j] = E[a_{l,j}^1|j] - E[a_{l,j}^0|j] \approx C_{H,l}H_l\Phi^{-1} \left( \frac{j-\frac{\pi}{8}}{j-\frac{\pi}{4}+1} \right),
\end{equation}
which is (\ref{eq:approx_delay}).\par

Simulation tests show that both (\ref{eq:approx_delay}) and (\ref{eq:first_m_avg_delay}) are very accurate in predicting holding delays.

\section{Proof of the divergence of cumulative means of a divergent increasing sequence}\label{apdx:abel_proof}
\setcounter{equation}{0} \renewcommand{\theequation}{B.\arabic{equation}}

For simplicity, assume a non-negative, divergent, and increasing sequence, $0\leq a_1 \leq a_2 \leq \ldots$, where $a_n\to\infty$ if $n\to\infty$. (We add the non-negativity condition because the delay terms in Section \ref{sec:bus_dynamics} are non negative. If this condition were to be removed, the claim is still true but the proof would be longer.) Define the cumulative mean sequence $b_n = \frac{1}{n}\sum_{i=1}^{n}{a_i}$. Apparently, $\{b_n\}$ is also non-negative and increasing.\par

\begin{proof}
	We prove $\{b_n\}$'s divergence by contradiction. Suppose $\exists M>0$, such that $b_{n}\leq M, \forall n$. Due to $\{a_n\}$'s divergence, $\exists N$ such that $a_N>2M$. Then $b_{2N}=\frac{1}{2N}\sum_{i=1}^{2N}{a_i}$ $=\frac{1}{2}(b_N + \frac{1}{N}\sum_{i=N+1}^{2N}{a_i})$  $>\frac{1}{2}(0+\frac{1}{N}\cdot N\cdot 2M)=M$, which contradicts what was assumed.
\end{proof}

\end{appendices}

\newpage
\begin{footnotesize}
\bibliographystyle{elsarticle-harv.bst}
\bibliography{literature.bib}
\end{footnotesize}
\end{document}